\documentclass[twocolumn]{svjour3}

\usepackage[dvips]{graphicx}
\usepackage{amsfonts}
\usepackage{amssymb}
\usepackage{amsmath}
\usepackage{bm}% bold math
\usepackage{hyperref}
\usepackage{mathrsfs}
\usepackage{color}
\renewcommand\floatsep{-18pt}
\renewcommand\intextsep{0pt}
\newcommand{\nb}{\nonumber}

\begin{document}
\title{Top-forms of Leading Singularities in Nonplanar Multi-loop Amplitudes}
\author{Baoyi Chen, Gang Chen,  Yeuk-Kwan E. Cheung, Ruofei Xie, Yuan Xin}
\institute{
G. Chen \at Department of Physics, Zhejiang Normal University, Jinhua, Zhejiang Province, China\\
Department of Physics, Nanjing University, 22 Hankou Road, Nanjing 210093, P. R. China
        %  \\
%             \emph{Present address:} of F. Author  %  if needed
           \and
           B.Y. Chen, Y.E. Cheung, R.F. Xie, Y. Xin \at
Department of Physics, Nanjing University, 22 Hankou Road, Nanjing 210093, P. R. China  
}
\maketitle
\begin{abstract}
The on-shell diagram is a very important tool in studying scattering amplitudes. In this paper we discuss the on-shell diagrams without external BCFW-bridges. We introduce an extra step of adding an auxiliary external momentum line. Then we can decompose the on-shell diagrams by removing external BCFW-bridges to a planar diagram whose top-form is well-known to all now. The top-form of the on-shell diagram with the auxiliary line can be obtained by adding the BCFW-bridges in an inverse order as discussed in our former paper \cite{Chen:2014ara}. To get the top-form of the original diagram,  the soft limit of the auxiliary line is needed. We obtain the evolution rule for the Grassmannian integral and the geometry constraint under the soft limit.  This completes  the top-form description of leading singularities in nonplanar scattering amplitudes of $\mathcal{N}=4$~Super Yang-Mills (SYM), which is valid for arbitrary higher-loops  and beyond the Maximally-Helicity-Violation (MHV) amplitudes.

\end{abstract}
%\pacs{11.15.Bt, 12.38.Bx, 11.25.Tq}
%\date{\today}

%\pacs{Valid PACS appear here}
% PACS, the Physics and Astronomy  Classification Scheme

\section{Introduction}
Bipartite diagrams and the associated  Grassmannian geometry~\cite{2006math......9764P,GeometryOnshell}
have recently found their way into the scattering amplitude studies.
An amazing discovery was to exploit them in computing  scattering amplitudes in $\mathcal{N}=4$ SYM 
theory~\cite{NimaGrass,NimaSMatrix,Paulos2014,BaiHe2014,Bargheer:2014mxa,Ferro2014gca,Franco2014csa,Elvang2014fja}.
Planar scattering amplitudes are represented by on-shell bipartite diagrams and expressed in ``top-form''
as contour integrations over the Grassmannian submanifolds.
Planar loop integrands  in $\mathcal{N}=4$ SYM have recently been constructed in~\cite{NimaGrass,arkani2012local} along  with the introduction of
the Grassmannian and on-shell method.
As a result, the ``$d$log'' form and the Yangian
symmetry~\cite{2010JHEP...04..085B,Broedel1403,Broedel2014,Beisert,Chicherin2014} of the scattering amplitudes are made manifest  in the planar limit.
It is  natural  to
extend the construction to non-planar scattering amplitudes~\cite{Chen:2014ara,Bern:2014kca,Arkani-Hamed:2014bca,Franco:2015rma},
and  theories of reduced (super) symmetries~\cite{Bidder2005N1,neitzke2014cluster,xie2012network}.

The leading singularities are represented in the
\emph{top-form} of Grassmannian integrals in which the integrands are comprised of  rational  functions of minors $\{\mathbf{R}(M_C)\}$ of the Grassmannian $C$ matrices. 
The top-form is elegant  in that the amplitude structures are simple and compact;  and the Yangian symmetry is manifest in the positive diffeomorphisms of positive Grassmannian 
geometry~\cite{NimaGrass}.
It is therefore crucial to express scattering amplitudes in top-form in order to explore its power to further  uncover hidden symmetries and dualities of the scattering amplitudes.
We present in this letter our successful construction of top-forms for non-planar scattering amplitudes. Our method applies to multi-loop, beyond-MHV leading singularities.

Recently, exciting progress in $\mathcal{N}=4$ SYM scattering amplitude 
computation (by  the on-shell method) is reported  by many 
research groups  in~\cite{Chen:2014ara,GeometryOnshell,Bern:2014kca,Arkani-Hamed:2014bca,BernNP12,du2014permutation,NimaSin2014,franco2015non,Johansson:2015ava}. Together we have made  a step forward in the computation of nonplanar 
$\mathcal{N}=4$ SYM scattering amplitudes, and hopefully  in the formulation   of AdS/CFT correspondence at  finite N. 

\section{BCFW-bridge decompositions of leading singularities}
\vspace{-10pt}
\makeatletter\def\@captype{figure}\makeatother
\hspace{0pt}\begin{minipage}{\linewidth}
%\centering
\includegraphics[width=0.8\columnwidth]{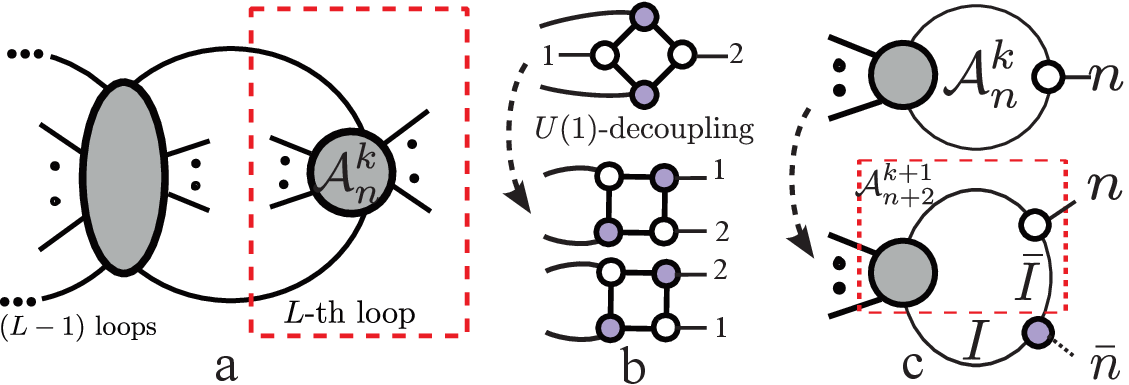}
\setlength{\abovecaptionskip}{-0pt}
\caption{(a)~Obtaining the $L$-th loop amplitude recursively. 
(b)~Utilising the $U(1)$ decoupling relation to turn a nonplanar diagram into a planar one. 
(c)~Introduction of  an auxiliary external momentum line to form the BCFW bridge.}
\label{fig:NLoopCut}
\vspace{10pt}
\end{minipage}

The aim of this work is to obtain a simple and compact analytical expression of leading singularities of scattering amplitudes, valid for arbitrary number of loops, beyond the planar limit. A general leading singularity can be represented by a reduced on-shell diagram. BCFW-bridge decomposition provides an efficient way of constructing on-shell diagrams in the planar limit. 
In non-planar cases, we can obtain the BCFW-bridge decomposition chain by extracting planar sub-diagrams and computing them  recursively~\cite{Chen:2014ara} as shown in Fig.~\ref{fig:NLoopCut}.  
For the sub-diagrams that are BCFW-decomposible, we follow the recipe presented in \cite{Chen:2014ara}. 
There exist, however, ``No Bridge'' (NB) diagrams which do not contain any BCFW bridges~\cite{Chen:2014ara,Arkani-Hamed:2014bca}.
We have presented a method in \cite{Chen:2014ara} to  transform some NB-diagrams, schematically depicted in Fig.~\ref{fig:NLoopCut}~(b),  by applying $U(1)$-decoupling relations \cite{KK}.
 
In this  work we present a general method applicable to  any NB-diagrams. The key  is to add an auxiliary external momentum line 
to  form an \textit{auxiliary BCFW bridge}, shown in Fig.~\ref{fig:NLoopCut}~(c). 
To regain the original NB diagram we  take the soft limit \cite{SoftLimitA,SoftLimitB,Volovich:2015yoa,Klose:2015xoa}, setting the auxiliary momentum to zero.  
This way  the BCFW-bridge decomposition chain of the reduced  on-shell diagrams beyond the planar limit can be obtained.

In the rest of the letter  we present a recipe for  constructing an analytical expression, {\it the top-form}, 
for a nonplanar leading singularity   using the BCFW-bridge 
decomposition chain \cite{Chen:2014ara} after  adding an  auxiliary external momentum line.

\section{Construction of the top-form}
The top-form of an on-shell diagram is obtained once the geometric constraints, $\Gamma$, and the integrand, $f(C)$, are determined. 
A non-planar leading singularity in the  form~\cite{NimaGrass}
\vspace{-5pt}
$$
\label{eq:TopForm}
\mathcal{T} = \oint_{\Gamma} {dC^{k\times n} \over \text{Vol}(GL(k))} {\delta^{k\times 4}(C\cdot\widetilde{\eta}) \over f(C)}\delta^{k\times2}(C\cdot\widetilde{\lambda})\delta^{2\times(n-k)}(\lambda\cdot C^\perp)~. \vspace{-5pt}
$$
requires to calculate $f(C)$ and $\Gamma$ under the BCFW shifts and take soft limit of the auxiliary BCFW bridges.  

Let us  study the BCFW shifts. 
The integrand, $f(C)$, must contain those poles equivalent to the constraints in $\Gamma$; otherwise the contour integration around $\Gamma$ will vanish.
Each BCFW bridge removes one pole in $f(C)$ by shifting a zero minor to be nonzero:  in general the poles in the integrand must change their forms  and the integrand changes its functional form accordingly.
To see this  we parametrize the constraint  matrix, $C$, using the BCFW parameter, $\alpha$.
In a BCFW shift, a column vector $X$ is shifted:
$
X\rightarrow \widehat{X}=X+\alpha Y,
$
with several minors of $f(C)$ become functions of $\alpha$.
After the shift, there exists at least one constraint $M_0(X)=0$ being shifted to  
$M_0(\widehat{X})=M_0(X)+\alpha R(Y)$ 
if there is a top-form. This is demonstrated in the following section.  Meanwhile the  factor $M_0(\widehat{X})$ should be present in the dominator to contribute a pole at  $\alpha=0$. 
In other words 
$
\alpha=M_0(\widehat{X})/R(Y)
$
is then a rational function of $\widehat{C}$
and can be subtracted from  other  shifted minors to obtain some shift-invariant  minors of $\widehat{C}$,
$
M_i(X)=M_i(\widehat{X}-\alpha Y).
$ 
In summary, attaching a  BCFW bridge, the integrand is, %
\vspace{-5pt}
\begin{equation}
\label{eq:TopformIntgrand}
  f(\widehat{C})
= M_0(\widehat{X}) \prod_i M_i(\widehat{X}-\alpha Y)
\times
\left(
   \begin{array}{c}
   \text{minors}\\ \text{without}~\alpha
   \end{array}
\right).
\vspace{-5pt}
\end{equation}

\subsection{An MHV example}
A six-point three-loop MHV example have been analyzed  in \cite{Arkani-Hamed:2014bca}. 
Here, for comparison, we provide our calculation by attaching the auxiliary BCFW-bridges. We attach an auxiliary external momentum  line, in Fig. \ref{mhveg},
\begin{figure}[htbp]
 \centering
 \includegraphics[width=0.15\textwidth]{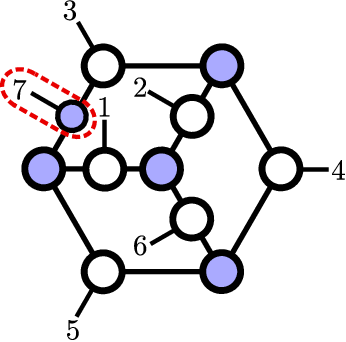}
 \caption{An MHV example}\label{mhveg}
 \label{mhveg}
\end{figure}
and form an auxiliary BCFW-bridge $(3,7)$. 
This on-shell diagram can be decomposed to identity as follows, 
$(1,7)\rightarrow(2,6)\rightarrow(3,5)\rightarrow(2,3)\rightarrow(3,4)\rightarrow(12)\rightarrow(2,3)\rightarrow(5,7) \rightarrow(3,7)$.  
Before adding Bridge-$(5,7)$, the on-shell diagram is still planar. The Grassmannian constraints and the top-form can be obtained directly from the permutation \cite{NimaGrass} shown in the first row of Tab.\ref{tab:MHVBD}. 
Adding  Bridge-$(5,7)$, the constraint $(7,1)^{1}\rightarrow (1,5,7)^2$, as shown in the second row of Tab.\ref{tab:MHVBD}. Here we use $(i_1,i_2,\cdots, i_m)^r$, where $m,r$ are positive integrals,  to denote the matrix of rank $r$ constructed by the columns $i_1,\cdots, i_m$ in the $C$ matrix. $(i_1,i_2,\cdots, i_m)^r$ also characterize the $(r-1)$-dimensional hyperplane in the $k$-dimensional projective space. 
Adding Bridge-$(3,7)$, the constraint $(1,5,7)^2$  disappears.
\begin{table}[htbp]
\vspace{10pt}
\caption{The evolution of the geometric constraints.}
\begin{center}
\begin{tabular}{|c|c|c|c|c|}\hline
~& $(7,1)^{1}$ & $(2,3,4)^{2}$ & $(4,5,6)^{2}$ & $(6,1,2)^{2}$ \\ \hline 
$(5,7)$ & $(1,5,7)^{2}$ & $(2,3,4)^{2}$ & $(4,5,6)^{2}$ & $(1,2,6)^{2}$ \\ \hline
$(3,7)$ & ~& $(2,3,4)^{2}$ & $(4,5,6)^{2}$ & $(1,2,6)^{2} $\\ \hline
\end{tabular}
\end{center}
\label{tab:MHVBD}
\end{table}
Before attaching Bridge-$(5,7)$, 
the top-form is 
$$
\frac{1}{(123)(234)(345)(456)(567)(671)(712)},
$$
where we use $(i_1i_2\cdots i_k)$ to denote the minor of the matrix which is formed by the  columns $i_1\cdots i_k$ in $C$.
When attaching $(5,7)$, the contour integral  around the pole $(712)=(\widehat{7}12)-\alpha (512)$ is replaced by 
$d\alpha\over\alpha$ with 
$\alpha = \frac{(\widehat{7}12)}{(512)}$. 
All the minors with column $7$ except $(712)$  are affected by the bridge 
$$
\begin{array}{ccccc}
(567) &\rightarrow& {(56(\widehat 7 5)\cap(12))\over (512)}&=&(56\widehat 7)\\ 
(671)&\rightarrow& {(6(\widehat 7 5)\cap(12)1)\over (512)} &=&\frac{(\widehat{7}15)(126)}{(512)},
\end{array}
$$
where the  $(\widehat 7 5)\cap(12)$ denotes the intersection point between the two lines characterized by $(\widehat 7 5)$ and $(12)$.
Then the top-form integrand  becomes
$$
\frac{(125)}{(123)(234)(345)(456)(567)(157)(126)(127)}.
$$
Similarly after attaching Bridge-$(3,7)$
the top-form integrand becomes 
$$
\frac{(135)^{2}}{(123)(234)(345)(456)(156)(357)(157)(126)(137)}.
$$
To obtain the  top-form of the original diagram, we parametrize the ${C^\prime}$ as 
$$
\left(
\begin{array}{cccccc|c}
 & &  &  C& &  & 0 \\
&  &  &  &  &  & 0 \\ \hline
0 & 0 & c_{3} & c_{4} & c_{5} & c_{6} & 1 \\
\end{array}
\right).
$$
Then we expand  all the minors in ${C^\prime}$ in terms of those in $C$
$$\begin{array}{ll}
(126)=(12) c_{6}& (234)=-c_{3}(24)+c_{4}(23)\\
(123)=(12)c_{3}&(456)=c_{4}(56)-c_{5}(46)\\
 (156)=-(16)c_{5}&(135)=-c_{3}(15)+c_5(13)\\
(137)=(13) &(345)=c_{3}(45)-c_{4}(35)+c_{5}(34)\\
(357)=(35)&(157)=(15) ~~ .
\end{array}$$
The top-form becomes 
\begin{eqnarray*}
&&\frac{(-c_{3}(15)+c_5(13))^{2}}{-(12)(35)(15)(12)(13)(16)c_3 c_6 c_5 (-c_{3}(24)+c_{4}(23))}\\
&\times&{1\over (c_{3}(45)-c_{4}(35)+c_{5}(34))(c_{4}(56)-c_{5}(46))}.
\end{eqnarray*}
The additional pole is characterised by $(126), (234), (456)$. The contour integration gives 
$c_{6}\rightarrow 0, c_{3}\rightarrow\frac{(23)}{(24)}c_{4}, c_{5}\rightarrow\frac{(56)}{(46)}c_{4}$ and 
\begin{eqnarray*}
\frac{[(56)(13)(24)-(23)(15)(46)]^{2}}{(23)(12)(24)(26)(45)(34)(16)(56)(35)(15)(13)(46)}
\end{eqnarray*}
consistent with the MHV example in \cite{Arkani-Hamed:2014bca}. 
This can be simplified further,  
\begin{eqnarray*}
-f_p(125364)+f_p(125463)-f_p(134265)\nonumber\\
+f_p(126543)-f_p(132465)-f_p(123564),
\end{eqnarray*}
with $f_p$ denoting the planar amplitudes of the corresponding orders.

\section{Construction of the top-from of NB diagram}
\label{sec:TFofNB}
Now we discuss  the new Grassmannian geometry structures in the NB diagrams~\cite{Chen:2014ara,Arkani-Hamed:2014bca}.
For the sub-diagram structure in Fig. \ref{fig:NLoopCut} (b), the top-form is obtained by imposing  $U(1)$-decoupling relation in~\cite{Chen:2014ara}; in this work we focus on the auxiliary BCFW bridges which is suitable for the general diagrams as shown in Fig. \ref{fig:NLoopCut} (c). 
The top-forms of those  diagrams  containing auxiliary BCFW bridges can be obtained using the above method. 
We discuss presently  how they return to the top-forms of the original NB diagrams upon  taking the soft limits.

The on-shell diagram $\mathcal{A}_{n+1}^{k+1}$ with one auxiliary line as shown in Fig.\ref{fig:NLoopCut} (c) can be written in two equivalent forms:
\begin{eqnarray}
\label{eq:RAmp}
\hspace{-20pt}&&\int \frac{d^2\lambda_{I} d^2\widetilde{\lambda}_{I}d^{4}\widetilde{\eta}_{I} }{vol(GL(1))}\frac{\mathcal{A}_{n+2}^{k+1}\langle I \bar I\rangle^3}{\left\langle \bar nI \right\rangle^{2}  \left\langle \bar{I} \bar n \right\rangle}\delta([\bar{n} I])\delta^{4}(\widetilde{\eta}_{\bar n}+{\langle\bar{I}\bar n\rangle\over \langle\bar I I\rangle}\widetilde{\eta}_{I}),~~~~\\
\label{eq:RTopForm}
\hspace{-20pt}&&\oint_{\bar{\Gamma}} {dC^{k\times n} dc_{k+1}\cdots dc_n \over \text{Vol}(GL(k))} {(1\cdots k )  \over f(C,c_i)}\nb\\
&&\times\delta^{k\times 4}(C\cdot\widetilde{\eta})\delta^{k\times2}(C\cdot\widetilde{\lambda})\delta^{2(n-k)}(\lambda\cdot {C^\prime}^\perp) \nb\\
&&\times\delta^{2}(\sum_{i=k+1}^{n} c_i \tilde\lambda_i+c_{\bar n} \tilde\lambda_{\bar n})\delta^{4}(\sum_{i=k+1}^{n} c_i \widetilde{\eta}_{i}+c_{\bar n} \widetilde{\eta}_{\bar n}) , 
\end{eqnarray}
where $c_{\bar n}=1$ and $\bar n$ denote the index of the auxiliary line. 
Eq.(\ref{eq:RAmp}) is obtained directly from 
Fig.\ref{fig:NLoopCut}(c) by integrating over the internal line $\bar I$.
Eq.(\ref{eq:RTopForm}) is a general top-form of 
$\mathcal{A}_{n+1}^{k+1}$,  where we choose a particular parametrization of the Grassmannian matrix ${C^\prime}$ as
$$
C^\prime=\left(
\begin{array}{cccccc|c}
   & & & & &  &0\\
 & & & C&  & &\vdots\\
   & & & & &  &0\\ \hline
0& \cdots&0  &c_{k+1}&\cdots& c_{n}&1
\end{array}
\right).
$$
Our method of adding the auxiliary line can be verified by comparing (\ref{eq:RAmp}) and (\ref{eq:RTopForm}). 
Noting that  $$\delta^{2}(\sum_{i=k+1}^{n} c_i \tilde\lambda_i+c_{\bar n} \tilde\lambda_{\bar n})=\delta(\sum_{i=k+1}^n c_i {[i1]\over [\bar n 1]}+1)\delta([\bar n  I]),$$
 (\ref{eq:RAmp}) and (\ref{eq:RTopForm}) can therefore be proved to  contain  the term $$\delta([\bar{n} I])\delta^{4}(\widetilde{\eta}_{\bar n}+{\langle\bar{I}\bar n\rangle\over \langle  \bar I I\rangle}\widetilde{\eta}_{I}),$$ which can be removed from  the overall  constraint delta function. 
The remaining  part of (\ref{eq:RAmp})  corresponds to the NB diagram $\mathcal{A}_n^k$ in the limit $\lambda_{\bar n}\rightarrow 0$. On the other hand, after taking the soft limit (\ref{eq:RTopForm}) yields 
 \begin{eqnarray}\label{eq:RRTopForm}
\mathcal{A}_n^k=\oint_{\bar{\Gamma}} {dC^{k\times n}dc_{k+1}\cdots dc_n \over \text{Vol}(GL(k))} {(1\cdots k)\delta^{k\times 4}(C\cdot\widetilde{\eta})\over f(C,c_i)} \nonumber\\
\times\delta^{k\times2}(C\cdot\widetilde{\lambda}) \delta^{2\times(n-k)}(\lambda\cdot {C}^\perp) \delta(\sum_{i=k+1}^n c_i {[i1]\over [\bar n 1]}+1)~.
 \end{eqnarray}
This is easily proven  by counting the degrees of freedom of the associated on-shell diagram in which only one element $c_f$ among $c_i$ is a free parameter. As we shall prove in the following section, that a given NB diagram has a top-form requires that  $c_i/c_r=\mathbf{R}(M_C)$. 
Using this relation, the integration $$\int{dc_r\over c_r}\delta(\sum_{i=k+1}^n c_i {[i1]\over [\bar n 1]}+1)$$ gives 1.
Finally we obtain the top-form of $\mathcal{A}_n^k$ by expanding the minors of ${C^\prime}$ into $ C$ minors  in the integrand.

\section{Rational top-forms and rational soft limit:}  \label{sec:rat}
Now we study on which kind of nonplanar on-shell diagrams  can  have  rational top-forms.  We  address  this question  by building  an equivalent relation between \textit{rational top-form} and \textit{rational soft limit}.
 If the soft limit of auxiliary line leads to additional constraints such that $c_i\over c_j$ is  a rational function of $C$-minors for all non-vanishing $c_i$ in the added  row of ${C^\prime}$, we call this soft limit \textit{a rational soft limit}.

When  the soft limit of the auxiliary line is a rational soft limit, then the NB diagram with auxiliary line has a rational top-form if and only if the original NB diagram  has a rational top-form. 

We fist consider the free parameters $\alpha$ in the top-form integrand $f(C)$  as shown in Eq. (\ref{eq:TopformIntgrand}). The $C$ matrix parameters $\alpha_C$ that can be expressed as  by $\mathbf{R}(M_C)$ is also of the form $\mathbf{R}(M_{C^\prime})$. 
The additional ${C^\prime}$  elements  $c_i$ are of the form 
$c_{r}\mathbf{R}(\alpha_C)$ 
(indicating a rational soft limit  for $c_{r}\neq 0$).  
Since $c_{r}$ is naturally $\mathbf{R}(M_{C^\prime})$, all free parameters in  ${C^\prime}$  are then rational functions of minors, i.e. the top-form is rational.  Inversely, given the linear auxiliary bridge and rational soft limit, any $C$ parameter denoted by $\mathbf{R}(M_{C^\prime})$ can be expanded as $\mathbf{R}(M_C)$ directly according to the procedure above. 

Then let us study the geometry constraints. Geometry constraints are linear relations among columns of the $C$ matrix. In fact, the total space is taken as the $(k-1)$-dimensional projective space. Each column labeled by the index of the external line can be mapped to a point in the projective space. For the diagram which can be constructed by attaching BCFW bridges, the constraints are all coplanarity constraints for the points of external legs \cite{Chen:2014ara}. For the NB diagrams, after attaching the auxiliary lines, the geometry constraints in $C'$ are still coplanarity constraints.  Hence we only need to discuss how the geometry constraints evolve in the rational soft limit.  

The simplest case is that the geometry constraints in $C'$ are all untangled. Then the coplanarity constraints are of the form 
$$(i_1, i_2,\cdots, i_m)^{m-1}.$$ If one of the indices, e. g. $i_m$,  in the above constraint denotes the auxiliary line, then the geometry constraint becomes  $(i_1, i_2,\cdots, i_{m-1})^{m-2}$ in the soft-limit. If none of the indices denotes the auxiliary line, then the geometry is invariant for $m<k$. Since $k\rightarrow k-1$ in the soft limit,  the geometry constraints for $m=k$ do not exist any more. 

In general the geometry constraints are still coplanarity constraints in the soft-limit. However, this is not obvious for the tangled cases. We will leave the explanation of the soft limit behavior for tangled geometry, e. g.  $((i_1 i_2\bigcap i_3 i_4),i_5,\cdots, i_m)^r$, to future work.  For now we focus on the algebraic form of these geometry constraints, which is enough to obtain the top-form. 

In a general case, the geometry constraints in $C'$ can be expanded as 
\begin{eqnarray}
\label{ExpandCon}
F'_1&=&\sum_{i=k+1}^{n}c_{i} F_1^{c_i}, ~~F'_2=\sum_{i=k+1}^{n}c_{i} F_1^{c_i},\nb\\
&\cdots&~~~~~~~~~~~F'_{n-k-1}=\sum_{i=k+1}^{n}c_{i} F_{n-k-1}^{c_i},\nb\\
&\cdots&~~~~~~~~~~~~~~~~~F'_{S}=\sum_{i=k+1}^{n}c_{i} F_{S}^{c_i},
\end{eqnarray}
where $S$ is an integer.  There are no higher order terms with respect to $c_i$'s. In fact if there are higher order terms, they can be factorized into linear polynomials either  with rational minors of $C$ as coefficients or with non-rational  minors. For the former case, one of the linear polynomials can be redefined as the geometry constraints. For the latter case, some $c_i/c_j$ are non-rational, which is beyond the scope of this paper. 

Among the constraints in Eq. \ref{ExpandCon},  we can choose arbitrary $n-k-1$ equations to solve for the $c_i/c_r$. For other equations, we can substitute the solutions of $c_i/c_r$ to get all the geometry constraints for $C$ after taking the soft limit.

\section{More Examples}
\subsection{An $\overline{\text{MHV}}$ example}
In this subsection we give an example of a different  
situation using an  $\overline{\text{MHV}}$ example (Fig.\ref{mhvbareg}) for illustration an auxiliary line connected  to a white vertex. 
Attaching an auxiliary line enables the  BCFW-decomposition to identity by the following chain: $ (7,3)\rightarrow(7,5)\rightarrow(3,4)\rightarrow(2,3)\rightarrow(4,5)\rightarrow(3,4)\rightarrow(1,3)\rightarrow(4,6) \rightarrow(3,7)$. 
 Before adding Bridge-$(7,5)$ the on-shell diagram is planar. 
 The Grassmannian constraints (the first row of Tab.\ref{tab:mhvbar}) and the top-form can be obtained directly from the permutation \cite{NimaGrass}:
$$
\frac{1}{(1234)(2345)(3456)(4567)(5671)(6712)(7123)}.
$$
 
\begin{figure}[htbp]
 \centering
 \includegraphics[width=0.15\textwidth]{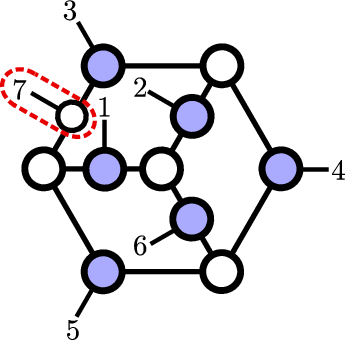}
 \caption{An $\overline{\text{MHV}}$ example}\label{mhvbareg}
 \label{mhvbareg}
\end{figure}

\begin{table}[htbp]
\vspace{10pt}
\caption{Geometry constraints evolution for $\overline{\text{MHV}}$ example}
\begin{center}
\begin{tabular}{|c|c|c|c|c|}\hline
& \tiny{$(2,(3,4,5)^{2},6)^{3} $}& $(3,4,5)^{2}$ &$ (5,6,7,1)^{3}$ & $(7,1,2,3)^{3}$ \\ \hline
$(7,5)$ & $(2,3,4,6)^{3}$ & $(3,4,5,7)^{3} $& $(5,6,7,1)^{3}$ & $(7,1,2,3)^{3}$ \\ \hline
$(7,3)$ & & $(3,4,5,7)^{3}$ & $(5,6,7,1)^{3}$ & $(7,1,2,3)_{3}$ \\ \hline
\end{tabular}
\end{center}
\label{tab:mhvbar}
\end{table}

The transformation of constraints after adding the bridges $(7,5)$ 
and $(7,3)$ is shown in the second and third rows of  
Tab.\ref{tab:mhvbar}.  The top-form of $\mathcal{A}_{7}^{4}$ becomes: 
$$
\frac{-(2467)^{2}}{(2347)(1246)(2456)(3457)(2346)(4567)(1567)(1267)(1237)}.
$$
Without loss of generality, we choose the first four columns of $\widehat{C}$ matrix as identity:
$$
\bordermatrix{
&\bf{e}_1&\bf{e}_2&\bf{e}_3&\bf{e}_4&&&\vec{c} \cr
&1 & 0 & 0 & 0 & * & * & c_{1}  \cr 
&0 & 1 & 0 & 0 & * & * & c_{2}  \cr 
&0 & 0 & 1 & 0 & * & * & c_{3} \cr 
&0 & 0 & 0 & 1 & * & * & c_{4} \cr 
}.
$$
Then the last column can be represented by these four columns as:
$$
\vec{c}=c_{1}{\bf{e}}_{1}+c_{2}{\bf{e}}_{2}+c_{3}{\bf{e}}_{3}+c_{4}{\bf{e}}_{4}.
$$
This way  we can rewrite the minor with column-7 as:
$$
\begin{array}{ll}
(1237)=c_{4}(1234)& (1567)=c_{2}(1256)+c_{3}(1356)\\
(1267)=c_{3}(1263) & (2467)=c_{1}(2461)+c_{3}(2463)\\
(2347)=c_{1}(1243)&  (3457)=-c_{1}(1345)-c_{2}(2345)\\
\multicolumn{2}{c}{(4567)=-c_{1}(1456)-c_{2}(2456)-c_{3}(3456)}
\end{array}.
$$

Consider the three poles of $\mathcal{A}_{7}^{4}$, since there is no constraint in the top-form of $\mathcal{A}_{6}^{4}$, we should integrate around all of the three poles  and remove three coefficients. 
At last, there is only one coefficient left and others can be represented by it:
$
c_{4}\rightarrow 0, 
c_{1}\rightarrow-\frac{(2345)}{(1345)}c_{2}, 
c_{3}\rightarrow-\frac{(1256)}{(1356)}c_{2}. 
$
The remaining  coefficient in the top-form is $\frac{dc_{2}}{c_{2}}$ and can be fixed by one of the columns in $\widehat{C}^{\perp}$ (noting that $\widehat{C}^{\perp}$ has one more column than $C^{\perp}$, which can be removed  directly). 
Finally, the top-form of $\mathcal{A}_{6}^{4}$ is:  
%\begin{scriptsize}
\begin{eqnarray*}
{1\over f_p}\frac{-[(2345)(1246)(1356)-(1256)(2346)(1345)]^{2}}{(1246)(2456)(1345)(2346)(1356)(1235)},
\end{eqnarray*}
%\end{scriptsize}
where $f_p=(1234)(2345)(3456)(4561)(5612)(6123)$.
This can also be simplified as 
\begin{eqnarray*}
&-&f_p(142356)-f_p(143265)+f_p(132456)  \nonumber\\
&-&f_p(132654)+f_p(123465)-f_p(123465).
\end{eqnarray*}

\subsection{An NMHV NB diagram example:}
\begin{figure}[htbp]
\begin{center}
\includegraphics[width=0.5\linewidth]{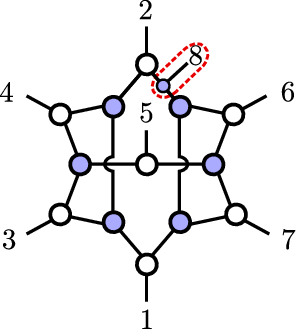}
\caption{A non-planar NMHV 4-loop leading singularity.}
\label{fig:NMHVeg}
\end{center}
\end{figure}

In this subsection  we present the details in the  calculation and simplification in the NMHV example (Fig.\ref{fig:NMHVeg}). 

Since this diagram cannot be decomposed by BCFW-bridges directly we introduce an auxiliary external momentum line the 
leg-{8}.   The diagram transforms to a planar one by removing the bridges $(2,8)$, $(6,8)$ and $(4,2)$. 
The total decomposition chain is 
$
(2,8)\rightarrow(6,8)\rightarrow(4,2)\rightarrow(1,2)\rightarrow(2,3)\rightarrow(2,4)\rightarrow(4,5)\rightarrow(4,6)\rightarrow(6,7)\rightarrow(1,6)\rightarrow(6,8).
$

Before adding Bridge-$(4,2)$  the planar diagram top-form is 
\begin{eqnarray}
1\over (1234)(2345)(3456)(4567)(5678)(6781)(7812)(8123)\nonumber
\end{eqnarray}
with constraints shown in the first row of Tab. \ref{tab:nmhveg}.
\begin{table}[htbp]
\vspace{10pt}
\caption{Geometry constraints evolution for NMHV example}
\begin{center}
\begin{tabular}{|c|c|c|c|c|}\hline
& $(1,2,3)^{2}$ & $(3,4,5)^{2}$ & $(5,6,7)^{2}$ & $(7,8,1)^{2}$ \\ \hline
$(4,2)$ & $(1,2,3,4)^{3}$ & $(3,4,5)^{2}$ & $(5,6,7)^{2}$ & $(7,8,1)^{2}$ \\ \hline
$(6,8)$ & $(1,2,3,4)^{3}$ & $(3,4,5)^{2}$ & $(5,6,7)^{2}$ & $(6,7,8,1)^{3} $\\ \hline
$(2,8)$ & $(1,2,3,4)^{3}$ & $(3,4,5)^{2}$ & $(5,6,7)^{2}$ & \\ \hline
\end{tabular}
\end{center}
\label{tab:nmhveg}
\end{table}
After attaching all the BCFW bridges, we obtain the top-form integrand
\begin{eqnarray}
&&{1\over (1234)(2345)(3456)(4567)(1567)(8672)(6781)}\nonumber \\
&\times&\frac{(1347)(6721)^{3}}{(7812)(3167)(1237)(1247)(1286)},
\end{eqnarray}
with geometry constraints as shown in the last row of  
Tab.\ref{tab:nmhveg}. 
Then we expand the rank-4 minors into  rank-3 minors 
$$
\begin{array}{ll}
(1234)=(123)c_{4} &(2345)=-(235)c_{4}+(234)c_{5}\\
(1237)=(123)c_{7}& (3167)=-(317)c_{6}+(316)c_{7}\\
(1286)=-(126) &(6721)=-(721)c_{6}+(621)c_{7}\\
(8672)=-(672) &(3456)=(356)c_{4}-(346)c_{5}\\
(6781)=-(671) & (1347)=-(137)c_{4}+(134)c_{7}\\
(7812)=(712) &(1247)=-(127)c_{4}+(124)c_{7} \\
\multicolumn{2}{c}{(1567)=(167)c_{5}-(157)c_{6}+(156)c_{7}}\\
\multicolumn{2}{c}{ (4567)=(467)c_{5}-(457)c_{6}+(456)c_{7}}.\\
\end{array}
$$
Solving all the additional constraints inherited  from the auxiliary line and the attached vertex,  we get 
$ c_{4}\rightarrow 0, c_{5}\rightarrow 0,  c_{7}\rightarrow \frac{(457)}{(456)}c_{6}$ and the final top-form intergrand
$${1\over f_p}\frac{(134)(357)[(457)(126)-(456)(127)]^{3}}{(124)(126)(135)(145)(267)(367)(457)^2},$$ where $f_p=(123)(234)(345)(456)(567)(671)(712)$.
Using the Pluker relations  the integrand $1\over f(C)$ is 
\begin{eqnarray}\label{Minor3}
&&{(125)\over (124)(126)(167)(235)(257)(143)(345)(567)}\nonumber\\
&+&{(125)\over (124)(127)(165)(235)(267)(143)(345)(567)}\nonumber\\
&+&{(125)\over (123)(126)(167)(234)(145)(257)(345)(567)}\nonumber\\
&+&{(125)\over (123)(127)(165)(234)(145)(267)(345)(567)}.
\end{eqnarray}
It is hard to simplify the form further  by using the Pl\"uker relations directly.  A simpler technique is to  para-metrize  the $C$ matrix as 
$$
\bordermatrix{%
&1&2&3&4&5&6&7 \cr 
 &*&* &*  &*  & 0&*  &*  \cr
&*& * & * &  *& 0 & * & * \cr
&c_1 & c_2 & c_{3} & 0& 1 & 0 & c_7 \cr
}
$$
and expand the three-column minors in the two-column minors. Then the first term in (\ref{Minor3}) can be written as
%\begin{tiny}
\begin{eqnarray}
&&{-(12)(36)\over (124)(126) c_7 (16)(23)(27)c_3(14)(34)(67)(36)}\nonumber\\
&=&{-1\over (124)(126)(27)c_3 (14) (34) (67) c_7 (36)}\nonumber\\
&+&{1\over (124)(126)c_7 (16) (37)c_3 (24) (34) (67) } \nonumber\\
&=&{-1\over (124)(126)(257) (134)(345) (567)(367)}\nonumber\\
&-&{1\over (124)(126)(167) (234) (357)(345)(567)} .
\end{eqnarray}
%\end{tiny}
Similarly we can rewrite the second term to forth term in~(\ref{Minor3}) as 
%\begin{small}
\begin{eqnarray}
(II) &=& {1\over (124)(127)(167) (346) (235)(345)(567)}\nonumber\\
&+& {1\over (124)(127)(134)(267)(356)(345)(567)}; \\
%\end{eqnarray}
%\begin{eqnarray}
(III)&=&{1\over (123)(126)(143) (467) (275)(345)(567)}\nonumber\\
&+&{1\over (123)(126)(243)(467)(175)(345)(567)}; \\
%\end{eqnarray}
%\begin{eqnarray}
(IV)&=&{1\over (123)(127)(143) (267) (456)(345)(567)}\nonumber\\
&+&{1\over (123)(127)(234)(167)(456)(345)(567)}.
\end{eqnarray}

\subsection{An on-shell diagram with two auxiliary lines}
\begin{figure}[htbp]
 \centering
 \includegraphics[width=0.3\textwidth]{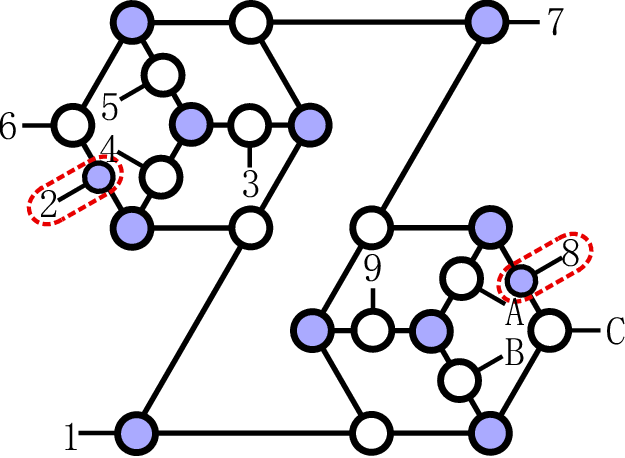}
 \caption{An $N^4$MHV example with two auxiliary lines}\label{NNNNmhveg}
\end{figure}

Our method can also deal with the on-shell diagrams  to which we need add more than one auxiliary lines. We consider the diagram in Fig. \ref{NNNNmhveg} where $k=4$ and $n=10$.  After attaching the BCFW bridges recursively 
\begin{eqnarray*}
(C8)&\rightarrow&(A8)\rightarrow(62)\rightarrow(42)\rightarrow(34)\rightarrow(23)\rightarrow(12)\rightarrow(45)\\
&\rightarrow&(34)\rightarrow(46)\rightarrow(37)\rightarrow(78)\rightarrow(27)\rightarrow(79) \rightarrow(7A)\\
&\rightarrow&(AB)\rightarrow(2A)\rightarrow(AC),
\end{eqnarray*}
we can get the integrand $f(C')$ of the top-form directly 
\begin{eqnarray}
&&\frac{1}{(56789A)(6789AC)(5679AB)(789ABC)}\nb\\
&\times&\frac{1}{(89ABC1)(BC1246)(ABC134)(9ABC13)}\nb\\
&\times&\frac{1}{(BC1236)(BC1234)(C12346)(BC1345)(123456)}\nb\\
&\times&\frac{(5679AC)^3(BC1346)^3}{(234567)(5678AC)(45679A)(345679)(56789C)}.
\label{toptwoline1}
\end{eqnarray}
The geometry constraints are also obtained immediately 
\begin{align}
&(3,4,5)^2,(9,A,B)^2,(8,A,B,C)^3,(2,4,5,6)^3,\nb\\
&(5,6,7,8,A,C,)^3, (1,4,5,6,7,9)^5, \nb\\
&(B,C,1,2,4,6)^3,(7,A,B,C,1,3)^5.~~~~~~
\label{GeometryC1}
\end{align}
We first take the soft limit of line-2. We use the gauge $GL(k+2)/GL(k+1)$ to set the $C''$ as 
$$
\bordermatrix{%
&1&2&3&4&5&6&7 &8 &9 &A &B&C \cr 
 &*&0 &*  &*  & *&*  &*  &* &* &*&*  &*\cr
  &*&0 &*  &*  & *&*  &*  &* &* &*&*  &*\cr
   &*&0 &*  &*  & *&*  &*  &* &* &*&*  &*\cr
 &*&0 &*  &*  & *&*  &*  &* &* &*&*  &*\cr
&*& 0& * &  *& * & * & * &*  &*&*&*  &*\cr
&0 & 1 & 0 & c_4& c_5 & c_6 & c_7 &c_8&c_9  &0&0 &0\cr
}.
$$
We pick up five linear independent equations with variables $c_i$, $i\in[4,9]$ from the constraints $(3,4,5)^2, (9,A,B)^2,  $ $(8,A,B,C)^3, (7,A,B,C,1,3)^5,(5,6,7,8,A,C)^3$ in Eq. (\ref{GeometryC1})  
\begin{eqnarray}
\label{LinearEqn1}
&&(35x_1x_2x_3)c_4-(34x_1x_2x_3)c_5=0\nb\\
&&(BAx_1x_2x_3)c_9=0\nb\\
&&(CABx_1x_2)c_8=0\\
&&(3ABC1)c_7=0\nb\\
&&(C678A)c_5+(5C78A)c_6+(56C8A)c_7+(567CA)c_8=0,\nb
\end{eqnarray}
where $x_i$ denotes an arbitrary index of the external line in each constraint equation. Obviously the chosen $x$ should leave the coefficients of $c_i$ in above equations non-vanishing.  We do not use the constraint containing line-2 in Eq. (\ref{GeometryC1})  since they do not generate any constraints for the $c_i$.  In the top-form, after taking the contour integration  of $dc_4\cdots dc_9$ around the pole generated by the above equations, we get
$c_7\rightarrow 0,\ c_8\rightarrow0,\ c_9\rightarrow0,\ c_4\rightarrow\frac{(34x_1x_2x_3)}{(35x_1x_2x_3)}c_5,\ c_6\rightarrow\frac{(678AC)}{(578AC)}c_5$  and leave with one integration $dc_5$. The  poles of the contour integration in Eq. (\ref{toptwoline1}) are reduced as the following 
\begin{align}
(5678AC)&=(578AC)(c_6-\frac{(678AC)}{(578AC)}c_5)\nb\\
(789ABC)&=-(89ABC)c_7\nb\\
(89ABC1)&=-(19ABC)c_8\nb\\
(9ABC13)&=-(13ABC)c_9\nb\\
(BC1345)&=-(135BC)(c_4-\frac{(BC134)}{(BC135)}c_5).
\end{align}
Other minors in Eq. (\ref{toptwoline1}) are reduced by substituting   the solutions of Eq. (\ref{LinearEqn1})
\begin{eqnarray*}
(123456)&=&(13456),~~
(234567)=-(34567),\\
(BC1246)&=&(146BC),
(BC1236)=(136BC),\\
(BC1234)&=&(134BC),
(C12346)=-(1346C),\\
(345679)&=&\frac{(34579)(678AC)}{(578AC)}c_5,\\
(56789C)&=&\frac{(5678C)(789AC)}{(578AC)}c_5,\\
(56789A)&=&\frac{(5678A)(789AC)}{(578AC)}c_5,\\
(6789AC)&=&\frac{(789AC)(678AC)}{(578AC)}c_5,\\
(5679AB)&=&\frac{(5679A)(79ABC)}{(579AC)}c_5,\\
(ABC134)&=&\frac{(13ABC)(3479A)}{(3579A)}c_5,\\
(5679AC)&=&-\frac{(567AC)(789AC)}{(578AC)}c_5,\\
(BC1346)&=&-\frac{(134BC)(378AC)(5678C)}{(578AC)(3578C)}c_5,\\
(45679A)&=&\frac{(4579A)(378AC)(5679A)}{(3579A)(578AC)}c_5.
\end{eqnarray*}

Then we get the top-form of the on-shell diagram in Fig. \ref{NNNNmhveg} after taking the soft limit of line-2
\begin{eqnarray}
&&\frac{1}{(13456)(1346C)(136BC)(13ABC)(146BC)(19ABC)}\nb\\
&\times&\frac{1}{(34567)(34579)(4578C)(5678A)(79ABC)(89ABC)}\nb\\
&\times&\frac{(134BC)(379AC)^2(5678C)^2(567AC)^3}{(579AC)(3578C)^2(5679A)^2(678AC)^2}.
\end{eqnarray}
The geometry constraints are 
\begin{eqnarray}
&&(3,4,5,6)^2,\ (9,A,B)^2,\ (8,A,B,C)^3,\  \nb\\
&&(B,C,1,4,6)^4,\ (1,5,6,7,9)^4.
\end{eqnarray}
We then take the soft-limit of line-8. We use the gauge $GL(k+1)/GL(k)$ to set the $C'$ as 
$$
\bordermatrix{%
&1&3&4&5&6&7 &8 &9 &A &B&C \cr 
 &*&*  &*  & *&*  &*  &0 &* &*&*  &*\cr
   &* &*  &*  & *&*  &*  &0 &* &*&*  &*\cr
 &*&*  &*  & *&*  &*  &0&* &*&*  &*\cr
&*& * &  *& * & * & * &0 &*&*&*  &*\cr
&c_1 & c_3 & 0 & 0& c_6 & 0 &1&0  &c_A&c_B &c_C\cr
}.
$$
Here we choose $c_4=c_5=c_7=c_9=0$. Using the constraints $(3,4,5)^2,\ (4,5,6)^2$,  $(9,A,B)^2, \ (B,C,1,4,6)^4,$ $ \ (1,5,6,7,9)^4$, we can find the poles at $c_3\rightarrow0,\ c_6\rightarrow0,\ c_1\rightarrow0,\ c_A\rightarrow\frac{(9Ax_1x_2)}{(9Bx_1x_2)}c_B,\ c_C\rightarrow\frac{(146C)}{(146B)}c_B$, which are shown in the integrand as
\begin{eqnarray*}
(13456)&=&(3456)c_1,\\
(34567)&=&-(3457)c_6,\\
(34579)&=&(4579)c_3,\\
(136BC)&=&(136B)(c_C-\frac{(136C)}{136B}c_B),\\
(79ABC)&=&(79BC)(c_A+(\frac{(146C)(79AB)}{(146B)(79BC)}-\frac{(79AC)}{(79BC)})c_B).
\end{eqnarray*}
Similar as the soft limit of line-2, the other minors are reduced as 
\begin{eqnarray*}
(3578C)&=&-(357C), (4578C)=-(457C),\\
(678AC)&=&(67AC), (89ABC)=(9ABC),\\
(5678A)&=&-(567A), (5678C)=-(567C),\\
{(134BC)\over(146BC)} &=&{(134B)\over(146B)},\\
(1346C)&=&\frac{(1346)(146C)}{(146B)}c_B,\\
(5679A)&=&\frac{(5679)(9A13)}{(9B13)}c_B,\\
\end{eqnarray*}
\begin{eqnarray*}
(13ABC)&=&-\frac{(1369)(13AB)(13BC)}{(139B)(136B)}c_B,\\
(19ABC)&=&-\frac{(1469)(14BC)(19AB)}{(146B)(149B)}c_B,\\
(579AC)&=&-\frac{(579A)(1469)(14BC)}{(149B)(146B)}c_B,\\
(379AC)&=&-\frac{(379A)(1469)(46BC)}{(469B)(146B)}c_B,\\
(567AC)&=&\left({(567A)(146C)\over(146B)}-{(567C)(9A13)\over(9B13)}\right)c_B.
\end{eqnarray*}
Then we get the integrand of the original on-shell diagram without auxiliary lines
\begin{eqnarray*}
&&\frac{1}{(1346)(1369)(3456)(3457)(4579)(5679)(579A)}\\
&\times&\frac{(134B)}{(146C)(37AB)(13BC)(19AB)(79BC)(9ABC)}\\
&\times&\frac{((367C)(379A)(146B)-(567A)(146C)(379B))^3}{(146B)^3(67AC)^2(357C)^2},
\end{eqnarray*}
with the geometry constraints $(3,4,5,6)^2, (9,A,B,C)^2$.

\section{Summary and Outlook}
We have obtained the top-form integrands for nonplanar leading singularities  by BCFW decompositions.  In the cases that one cannot attach a BCFW-bridge we  add an auxiliary external momentum line judiciously  to enable the application of the chain of BCFW decompositions and take the soft limit on the auxiliary momentum line to recover the original diagrams. 
This combination of  strategies is efficient on computing the leading singularity of nonplanar diagrams of arbitrary loops.  
We  have also  classified  nonplanar on-shell diagrams 
according to whether they possess rational top-forms, 
and proved its equivalence to linear BCFW bridges (and rational soft limit for diagrams with no external BCFW bridges). 
With the  chain of BCFW-bridge decompositions obtained  the
rational top-forms of the nonplanar on-shell diagrams can be derived  in a straightforward way.  
This method applies to  leading singularities of 
nonplanar multi-loop amplitudes beyond MHV. 

An immediate question is whether all on-shell diagrams representing nonplanar leading singularities belong to 
this class, so that all leading singularities can be 
expressed in the rational top-forms. 

Top-form, being simple and compact,  is a usefull tool to uncover  hidden symmetries  (e.g. generalized Yangian symmetry beyond 
planarity~\cite{du2014permutation})
which are otherwise highly tangled in nonplanar
leading singularities.
When combined with the generalized unitarity cuts, top-form holds promise in constructing the integrals as well as
revealing the symmetries and dualities  
of loop-level scattering amplitudes.

Mathematically  our method of performing the BCFW decompositions is related to the toric geometry arising in the characterization of Matroid Stratification.
Further exploration on the relationship between BCFW decompositions  and Matroid Stratification  will also shed light on the geometry of underlying Grassmannian  manifolds.

\section{Acknowledgments}
G. Chen thanks Nima Arkani-Hamed, Tianheng Wang for helpful discussion and useful comments. We
thank Peizhi Du, Shuyi Li and Hanqing Liu for constructive discussion. Yuan Xin
thanks Bo Feng for introducing the background on the recent developments of scattering
amplitude. This research project has been supported by the Fundamental Research
Funds for the Central Universities under contract 020414340080, the Youth Foundation
of China under contract 11405084.  This research project has been supported in parts by the NSF China under Contract No.~11775110,  No.~1169-0034. We also acknowledge the European Union's Horizon 2020 Research and Innovation~(RISE) programm under the Marie Sk\'lodowska-Curie grant agreement No.-644121, and  the Priority Academic Program Development for Jiangsu Higher Education Institutions (PAPD).
%\end{small}
\appendix
\label{app}

\end{document}